\documentclass[epj,referee]{svjour}
\usepackage[dvipdfm]{graphicx}
\usepackage{amsmath}
\usepackage{amssymb}
\newcommand{\upd}{\mathrm{d}}
\renewcommand{\vec}[1]{\mbox{\boldmath $#1$}}
\newcommand{\tensor}[1]{\mbox{\boldmath $#1$}}

\begin{document}
\title{
Simulating (electro)hydrodynamic effects in colloidal
dispersions: smoothed profile method
}
\author{Yasuya Nakayama\inst{1}
\fnmsep
\thanks{\email{ynakayama@chem-eng.kyushu-u.ac.jp}}%
 \and Kang Kim\inst{2} \and
Ryoichi Yamamoto\inst{3,4}
}                     
\institute{
Department of Chemical Engineering, Kyushu University,
  Fukuoka 819-0395, Japan
  \and
Department of Computational Molecular Science, Institute for Molecular
Science, Myodaiji, Okazaki, Aichi 444-8585, Japan 
  \and
Department of Chemical Engineering, Kyoto University,
  Kyoto 615-8510, Japan
\and
CREST, Japan Science and Technology Agency - 4-1-8 Honcho
  Kawaguchi, Saitama 332-0012, Japan
}
\date{Received: date / Revised version: date}
\abstract{
Previously, we have proposed a direct simulation scheme for colloidal dispersions in
a Newtonian solvent~[Phys.Rev.E {\bfseries 71},036707 (2005)].  An
improved formulation called the ''Smoothed Profile~(SP) method'' is
presented here in which simultaneous time-marching is used for the host fluid and
colloids.  The SP method is a direct numerical simulation of
particulate flows and provides a coupling scheme between the continuum fluid
dynamics and rigid-body dynamics through utilization of a smoothed profile for the colloidal
particles.  Moreover, the improved formulation includes an extension to
incorporate multi-component fluids, allowing systems such as charged colloids in
electrolyte solutions to be studied.  The dynamics of the colloidal dispersions are solved with the same computational cost as required for solving non-particulate flows.
Numerical results which assess the hydrodynamic interactions of colloidal
dispersions are presented to validate the SP method.  The SP method is not
restricted to particular constitutive models of the host
fluids and can hence be applied to colloidal dispersions in
complex fluids.
} 
\maketitle
\section{Introduction}
Interparticle interactions in colloidal dispersions mainly consist of
thermodynamic potential interactions and hydrodynamic
interactions~\cite{happel83:_low_reynol,kim91:_microh,russel89:_colloid_disper}.
Whereas the former interactions occur in both static and dynamic
situations, the latter occur solely in dynamic situations. Although
thermodynamic interactions have been studied extensively and summarized
as a concept of the effective interaction~\cite{likos01:_effec}, the
nature of dynamic interactions is poorly understood.  Since the
hydrodynamic interaction is essentially a long-range, many-body effect,
it is extremely difficult to study its role using analytical methods
alone. Numerical simulations can aid the investigation of the
fundamental role of hydrodynamic interactions in colloidal dynamics.

In recent decades, various simulations for particulate flow have
been developed for most simple situations in which the host fluid is
Newtonian~\cite{brady88:_stokes_dynam,malevanets99:_mesos,tanaka00:_simul_method_colloid_suspen_hydrod_inter,kajishima01:_turbul_struc_partic_laden_flow,hu01:_direc_numer_simul_fluid_solid,glowinski01:_fictit_domain_approac_direc_numer,ladd01:_lattic_boltz,padding04:_hydrod_brown_fluct_sedim_suspen,cates04:_simul_boltz}.
However, these schemes can not necessarily be applied to problems with
non-Newtonian host fluids or solvents with internal microstructures,
which are practically more important cases.  Hydrodynamic simulations of
ions and/or charged colloids have been proposed by making use of some of
the above
schemes~\cite{kim05:_elect,yamaue05:_multi_phase_dynam_progr_manual,lobaskin04:_elect,chatterji05:_combin_lattic_boltz,capuani04:_discr,capuani06:_lattic_boltz}.
Nonetheless, the tractability and/or physical validity of their modeling
remain controversial.
In Ref.\cite{kim05:_elect}, in solving the motion of the ionic solutes,
the solvent hydrodynamic interaction was incorporated by
Rotne--Pragar--Yamakawa-type mobility tensor, which accounted for a
long-range part of pair interaction.  This scheme misses many-body
and/or near-field hydrodynamic interactions between the solutes and
       surfaces, which
can have an effects on the behavior of ions near surface.
In Ref.\cite{yamaue05:_multi_phase_dynam_progr_manual}, the
computational mesh was arranged to express round shape of a colloidal
particle and the boundary value problem of the solvent continuum
equations was solved based on the irregular mesh.  For applying this
scheme to many-body system, inaccessible computational resources is
inevitable.
In
Refs.\cite{lobaskin04:_elect,chatterji05:_combin_lattic_boltz,capuani04:_discr,capuani06:_lattic_boltz},
they adopted the lattice Boltzmann~(LB) method for explicitly solving
the solvent hydrodynamics.  Although the LB method has many advantages
for solving large systems, compared to the standard discretization
schemes, the applicability to various constitutive equations for the
complex fluids other from the Newtonian fluids is unveiled.
Concerning the interaction between colloidal particles and the solvent,
authors in
Refs~\cite{lobaskin04:_elect,chatterji05:_combin_lattic_boltz}, arranged
finite interaction points on the surface of the colloids on which a
frictional coupling was assumed.  In this coupling scheme, friction
parameters and the number of the interaction points were determined
phenomenologically and theoretical basis to determine the parameters in
general systems was not clearly stated.

In this article, we propose a direct simulation scheme for colloidal
dispersions which is applicable to most constitutive models of the host
fluids. We call it the ''Smoothed Profile~(SP) method'' since the
original sharp interface between the colloids and solvent is replaced
with an effective smoothed interface with a finite
thickness~\cite{tanaka00:_simul_method_colloid_suspen_hydrod_inter,nakayama05:_simul_method_resol_hydrod_inter_colloid_disper}.
We formulate a computational method to couple the particle dynamics and
hydrodynamics of the solvent.  A fixed grid is used in both the solvent
and the particle domains.  Introduction of a smoothed profile makes it
possible to realize stable and efficient implementation of our scheme.
The numerical implementation for Newtonian solvents and electrolyte
solutions as specific examples of multi-component fluids is
outlined. Various test cases which verify the SP method and assess
the hydrodynamic interactions are presented.

\section{Dynamics of a multi-component solvent and colloids}
\subsection{Hydrodynamics of multi-component fluids}
We first give a brief description of multi-component fluid equations,
looking at electrolyte solutions as a specific application.
Consider \(N\) (possibly ionic) solute species that satisfy the law of conservation for every concentration,
\(C_{\alpha}\), of the \(\alpha\)th species:
\begin{align}
\partial_{t}C_{\alpha} + \nabla \cdot C_{\alpha}\vec{v}_{\alpha} +
\nabla\cdot\vec{g}_{\alpha} &= 0,
\end{align}
where \(\vec{v}_{\alpha}\) is the velocity of the \(\alpha\)th solute and
\(\vec{g}_{\alpha}\) is a random current.  Since the inertial time
scales of the solute molecules are extremely small, the velocity of the
\(\alpha\)th solute can be decomposed into the velocity of the solvent \(\vec{v}\)
and the diffusive current arising from the chemical potential gradient
\(\nabla \mu_{\alpha}\) as: 
\begin{align}
\vec{v}_{\alpha} &=\vec{v} -
\Gamma_{\alpha}\nabla \mu_{\alpha},
\end{align}
 where \(\Gamma_{\alpha}k_{B}T\) is the diffusivity of the \(\alpha\)th ion,
\(k_{B}\) is the Boltzmann constant and \(T\) is the temperature. The random
current should satisfy the following fluctuation-dissipation relation~\cite{landau59:_fluid}:
\begin{align}
\langle g_{\alpha,i}(\vec{x},t)g_{\beta,j}(\vec{x}',t')\rangle 
&=
2(k_{B}T)^{2}\Gamma_{\alpha}
\delta_{\alpha\beta}\delta_{ij}\delta(\vec{x}-\vec{x}')\delta(t-t')
.
\end{align}
Here, for the sake of simplicity, we do not deal with the cross
diffusion of different solutes.
The conservation of momentum implies the velocity of solvent follows the
Navier--Stokes equation of incompressible flow with the source term from
solutes:
\begin{align}
\nabla \cdot \vec{v} &= 0,
\\
\rho \left(\partial_{t}+\vec{v}\cdot\nabla \right)\vec{v} &= -\nabla p
+ \eta \nabla^{2}\vec{v} -\sum_{\alpha}C_{\alpha}\nabla \mu_{\alpha} 
+\nabla\cdot \tensor{s},
\end{align}
where \(\rho\) is the total mass density of the fluid, \(p\) is the
pressure, \(\eta\) is the shear viscosity of the fluid, and \(\tensor{s}\) is a
random stress satisfying the fluctuation-dissipation relation~\cite{landau59:_fluid}:
\begin{align}
\langle
s_{ik}(\vec{x},t) s_{jl}(\vec{x}',t') \rangle 
\&=
2k_{B}T\eta
(\delta_{ij}\delta_{kl}+\delta_{il}\delta_{kj})\delta(\vec{x}-\vec{x}')\delta(t-t')
.
\end{align}

The above set of equations is closed when a set of chemical potentials
\(\{\mu_{\alpha}\}\) is given, and describes the dynamics of a
multi-component fluid. For the specific application of an electrolyte solution, we consider the Poisson--Nernst--Planck equation for the chemical potential:
\begin{align}
\mu_{\alpha}(\{C_{1},\ldots,C_{N}\}) &= k_{B}T \log C_{\alpha} +
Z_{\alpha} e (\Phi - \vec{E}^{ext}\cdot \vec{x}),
\\
\epsilon
\nabla^{2}\Phi &= -\rho_{e},
\end{align}
This equation describes the Poisson--Boltzmann distribution for ions at
equilibrium, 
where \(Z_{\alpha}\) is the valence of the \(\alpha\)th ion, \(e\) is the
elementary charge, \(\Phi\) is the electrostatic potential,
\(\vec{E}^{ext}\) is the external field, \(\epsilon\) is the dielectric
constant of the fluid, and \(\rho_{e}\) is the charge density field. This
set of equations corresponds to the electrokinetic equations which appear in
standard textbooks~\cite{russel89:_colloid_disper}.

\subsection{Colloids in electrolyte solutions}
The colloid dynamics are maintained by the force exerted by the solvent.
Consider monodisperse spherical colloids with a radius \(a\), a mass
\(M_{p}\), and a moment of inertia \(\tensor{I}_{p}\). Momentum
conservation between the fluid and the \(i\)th colloid implies the following
hydrodynamic force and torque:
\begin{align}
\vec{F}_{i}^{H} &=\int
(\upd\vec{S}_{i}\cdot\tensor{\sigma}),~~
\vec{N}_{i}^{H} =
\int \left( \vec{x}-\vec{R}_{i} \right)\times (\upd\vec{S}_{i}\cdot
\tensor{\sigma}),
\end{align}
 where \(\vec{R}_{i}\) is the center of mass,
\(\int \upd\vec{S}_{i}(\dots)\) indicates the surface integral over the \(i\)th
colloid, and \(\tensor{\sigma}\) is the stress tensor of the fluid.
In terms of the electrokinetic equations, the stress reads as
\begin{align}
\tensor{\sigma} &= -p\tensor{I} + \tensor{\sigma}' +
\tensor{\sigma}^{st} + \tensor{s},
\end{align}
in which \(
\tensor{\sigma}' = \eta \left(\nabla \vec{v}+\left(\nabla
\vec{v}\right)^{T}\right)\)
 is the dissipative stress, and
\(
\tensor{\sigma}^{st} = \epsilon \left\{\vec{E}\vec{E}
 -(\left|\vec{E}\right|^{2}/2)\tensor{I}\right\}
\)
 is the Maxwell stress
for the electric field $\vec{E}=-\nabla \Phi + \vec{E}^{ext}$, where $\tensor{I}$ is the
unit tensor.
The evolution of the colloids follows Newton's equations:
\begin{align}
\dot{\vec{R}}_{i} &= \vec{V}_{i}, 
\\
M_{p}\dot{\vec{V}}_{i} &=
  \vec{F}_{i}^{H} + \vec{F}_{i}^{c}+ \vec{F}_{i}^{ext},
\\ 
 \tensor{I}_{p}\cdot\dot{\vec{\Omega}}_{i} &=
  \vec{N}_{i}^{H}+\vec{N}_{i}^{ext},
\end{align}
where \(\vec{F}_{i}^{ext}\) and \(\vec{N}_{i}^{ext}\) are the external
force and torque, respectively, and \(\vec{F}_{i}^{c}\) is the force
arising from the core potential of the particles, which prevents the colloids
overlapping. Hereinafter, the soft-core potential of the truncated
Lennard--Jones potential is adopted for \(\vec{F}_{i}^{c}\).
For the charged-colloid system specifically, the buoyancy is included in the term \(\vec{F}_{i}^{ext}\), and the external field on the colloids is accounted for by
\(\vec{F}_{i}^{H}\).

Having colloidal particles with a finite volume provides the relevant
boundary conditions to the hydrodynamic equations.  For the solvent
velocity, a no-slip condition is assigned, such that \(\vec{v} =
\vec{V}_{i}+\vec{\Omega}_{i}\times \vec{r}_{i}\) with
\(\vec{r}_{i}=\vec{x}-\vec{R}_{i}\) for the \(i\)th colloid.  For the
concentration field, a no-penetration condition is assumed, giving
\(\vec{n}\cdot \nabla \mu_{\alpha} = 0\), where \(\vec{n}\) represent
the unit normal to the surface of the colloids.  Coupling the
hydrodynamics of the solvent with the dynamics of the colloids defines
the moving-boundary-condition problem above.  The usual numerical
techniques of partial differential equations are hopeless in dealing
with the dynamical evolution of many colloids since the sharp interface
at the surface of the colloids moves and henceforth the mesh points at
which the boundary condition is assigned vary with each discrete time
step. The moving boundary condition leads to huge computational costs.

In contrast, the SP method formulates an efficient scheme for this kind
of moving-boundary-condition problem, and incurs the same level of
computational cost as is required for solving a uniform fluid.
The typical computational costs for the fluid and particles are assumed
to scale to their degrees of freedom.  
In the SP method, regular mesh, but not body-fitted mesh, can be used,
which indicates that the inclusion of the dispersed particle phase does
not induce the increase of the grids.
In \(d\)-dimensional system, we
assume that discretized space contains \(N^{d}\) grids and the variables
of the fluid phase are linked to the grids. \(N_{p}\) particles are
dispersed in the system and their volume scale to \(N_{p}a^{d}\). The
number of the particles are at most \(N_{p}a^{d}< (N\Delta x)^{d}\)
where \(\Delta x\) is the lattice spacing, which results in
\(N_{p}/N^{d} <\left(\Delta x/a\right)^{d}\). This inequality indicates
that the computational costs for the dispersed phase tracking is at
largest \(\left(\Delta x/a\right)^{d}\) times for the fluids. In typical
case of \(a/\Delta x = 5\), the fraction of the computational costs for
the particles roughly estimated less than several percent,
thus 
most of the computation should be for solving the uniform fluid.

\section{Computational algorithm}
In the SP method, quantities are defined over the entire domain, which
consists of the fluid domain and the particle domain. To designate the
particle domain, we introduce a concentration field for the colloids, given as
\(\phi(\vec{x},t) = \sum_{i=1}^{N_{p}}\phi_{i}(\vec{x},t)\),
where \(\phi_{i}\in [0,1]\) is the profile field of the \(i\)th particle, which is unity at the particle domain, zero at the fluid domain, and which has a
continuous diffuse interface of finite thickness \(\xi\) at the interface domain.
With the field \(\phi\), the total velocity field and concentration fields
of the solutes are defined as
\begin{align}
 \vec{v} &= (1-\phi)\vec{v}_{f} +\phi \vec{v}_{p},
\\
C_{\alpha} &= (1-\phi)C_{\alpha}^{*},
\end{align}
where \((1-\phi)\vec{v}_{f}\) represents the velocity field of the fluid,
and \( \phi\vec{v}_{p}(\vec{x},t) = \sum_{i=1}^{N_{p}}\phi_{i}(\vec{x},t)
\left[ \vec{V}_{i}(t) + \vec{\Omega}_{i}(t) \times \vec{r}_{i}(t) \right] \) is the velocity field of the colloids.
The auxiliary concentration field \(C_{\alpha}^{*}\)
is introduced, which can have a
finite value in the particle domain, whereas $C_{\alpha}$, the physical
concentration field, is forced to be zero through multiplication by
\((1-\phi)\).

The advection of \(\phi\) is solved via
\(\dot{\vec{R}}_{i}=\vec{V}_{i}\) and by mapping
\(\{\vec{R}_{1},\ldots,\vec{R}_{N_{p}}\}\) to \(\phi\). Henceforth, the
volume of the fluid and/or the solid is strictly conserved and no
numerical diffusion of \(\phi\) occurs.  In the SP method, the
fundamental field variables to be solved are taken as the total velocity
\(\vec{v}\) rather than \(\vec{v}_{f}\), and \(C_{\alpha}^{*}\) rather
than \(C_{\alpha}\).  This choice of variables yields great benefits in
terms of allowing efficient and stable time evolution.  The evolution
equation of \(\vec{v}\) is derived based on momentum conservation
between the fluid and the particles, and the rigidity of the particle
velocity field \(\vec{v}_{p}\).
For the solute concentration, \(C_{\alpha}\) and \(C_{\alpha}^{*}\)
differ in terms of whether they exhibit abrupt variation at the
interface of the colloids or not: especially for \(\xi\to 0\),
\(C_{\alpha}\) has discontinuity at the solid-fluid interface, but
\(C_{\alpha}^{*}\) should not.
Since \(C_{\alpha}\) exhibits abrupt variation in the spacial scale of
\(\xi\), in order to solve its advection, it is necessary to stabilize
the evolution over time and to prevent numerical diffusion and
penetration of \(C_{\alpha}\) into the particle domain.
Compared with \(C_{\alpha}\), auxiliary concentration \(C_{\alpha}^{*}\)
is independent of the abrupt variation arising from \((1-\phi)\), and
finite value of \(C_{\alpha}^{*}\) in the particle domain is allowed.
Therefore, it is numerically much easier to solve the advection equation
of \(C_{\alpha}^{*}\) than \(C_{\alpha}\).

\subsection{Discretization in time}
The time-discretized evolution equations are derived as follows.  To simplify the presentation, we neglect random currents.  As an
initial condition at the \(n\)th discretized time step, the position,
velocity, and angular velocity of the colloids,
\(\{\vec{R}_{i}^{n},\vec{V}_{i}^{n},\vec{\Omega}_{i}^{n}\}\)
\((i=1,\ldots,N_{p})\), are mapped to \(\phi^{n}$ and $\phi^{n}
\vec{v}_{p}^{n}\) and the following conditions are set:
\(\vec{v}^{n}=(1-\phi^{n})\vec{v}_{f}+\phi^{n}\vec{v}_{p}^{n}\), satisfying
the incompressibility condition on the entire domain, \(\nabla\cdot
\vec{v}^{n}=0\), and \(C_{\alpha}^{*,n}\), satisfying the charge neutrality
condition, \(\int \upd\vec{x}(1-\phi^{n})\sum_{\alpha}Z_{\alpha}e
C_{\alpha}^{*,n} + \int \upd\vec{x}\left|\nabla
\phi\right|e\sigma_{e}=0\), where \(\left|\nabla
\phi_{i}\right|e\sigma_{e}\) represents the surface charge distribution
of the colloids.  The current for the auxiliary concentration
field is defined as
\begin{align}
 C_{\alpha}^{*}
\vec{v}_{\alpha}
 &=
 C_{\alpha}^{*}
\vec{v} +
\left(\tensor{I}-\vec{n}\vec{n}\right)\cdot
C_{\alpha}^{*}
\left(
-\Gamma_{\alpha}\nabla \mu_{\alpha}
\right),
\end{align}
where \(\vec{n}(\vec{x},t)\) is the unit surface-normal vector field
which is defined on the interface domain with a finite thickness
\(\xi\).
In this definition of the current, the no-penetration condition
is directly assigned.
The auxiliary concentrations are advected by this current as
\begin{align}
C_{\alpha}^{*,n+1} &= 
 C_{\alpha}^{*,n}
- \int_{t_{n}}^{t_{n}+h}\upd s\nabla \cdot C_{\alpha}^{*}\vec{v}_{\alpha},
\end{align}
where \(h\) is a time increment, and \(t_{n}=nh\) is the \(n\)th
discretized time.
The total velocity field is updated using a fractional step approach.
First, the advection and the hydrodynamic viscous stress are solved, 
\begin{align}
 \vec{v}^{*} 
&=
\vec{v}^{n} 
+
\int_{t_{n}}^{t_{n}+h}
\upd s
\nabla 
\cdot 
\left[
\frac{1}{\rho}\left(
-p\tensor{I} +\tensor{\sigma}'
\right)
-\vec{v}\vec{v} \right],
\\
\vec{R}_{i}^{n+1} &= \vec{R}_{i}^{n} + \int_{t_{n}}^{t_{n}+h}\upd s \vec{V}_{i},
\end{align}
with 
the incompressibility condition,
\(\nabla\cdot\vec{v}^{*}= 0\).
Along with the advection of the total velocity, 
the particle position is updated using the particle velocity.
The electrostatic potential for the updated particle configuration is
determined by solving the following Poisson equation, 
\begin{align}
\nabla^{2}\Phi^{n+1} 
&=
-\rho_{e}^{n+1}/\epsilon,
\end{align}
with the charge density field,
$
\rho_{e}^{n+1} 
=
(1-\phi^{n+1})
\sum_{\alpha}Z_{\alpha}eC_{\alpha}^{*,n+1} 
+\left|
\nabla
\phi^{n+1}
\right|e\sigma_{e}$.
The momentum change as a result of the electrostatic field is solved as
\begin{align}
\vec{v}^{**} &= \vec{v}^{*} -h\rho_{e}^{n+1}\nabla \Phi^{n+1}.
\end{align}
At this point, the momentum conservation is entirely solved for the total
velocity field. The rest of the updating procedure applies to the rigidity
constraint on the particle velocity field.

The hydrodynamic force and torque on the colloids 
exerted by the fluid
are derived by assuming momentum conservation between the colloids and the
fluid. The time-integrated hydrodynamic force and torque over a period
\(h\) are equal to the momentum change over the particle domain:
\begin{align}
\left[
 \int_{t_{n}}^{t_{n}+h}
\!\!\!\!\!\!\!\!\!
\upd s\vec{F}_{i}^{H}(s)
\right]
 &= 
\int \upd\vec{x}
\rho\phi_{i}^{n+1}\left(
\vec{v}^{**}-\vec{v}_{p}^{n}
\right),
\label{eq:hydro_force}
\\
\left[
\int_{t_{n}}^{t_{n}+h}
\!\!\!\!\!\!\!\!\!
\upd s
\vec{N}_{i}^{H}(s)
\right]
  &= 
  \int 
\upd\vec{x}
  \left[
   \vec{r}_{i}^{n+1}
\times
 \rho\phi_{i}^{n+1}\left(
\vec{v}^{**}-\vec{v}_{p}^{n}
\right)
\right].
\label{eq:hydro_torque}
\end{align}
With this and other forces on the colloids,
the particle velocity and angular velocity are updated as
\begin{align}
 \vec{V}_{i}^{n+1} &= \vec{V}_{i}^{n} 
+
M_{p}^{-1}
\left[
\int_{t_{n}}^{t_{n}+h} 
\!\!\!\!\!
\!\!\!
\upd s 
\vec{F}_{i}^{H}
\right]
+
M_{p}^{-1}
\int_{t_{n}}^{t_{n}+h} 
\!\!\!\!\!
\!\!\!
\upd s 
\left(
\vec{F}_{i}^{c}
+
\vec{F}_{i}^{ext}
\right),
\\
 \vec{\Omega}_{i}^{n+1} &= \vec{\Omega}_{i}^{n} 
+\tensor{I}_{p}^{-1}
\cdot
\left[
\int_{t_{n}}^{t_{n}+h} 
\!\!\!\!\!
\!\!\!
\upd s 
\vec{N}_{i}^{H}
\right]
+
\tensor{I}_{p}^{-1}
\cdot
\int_{t_{n}}^{t_{n}+h} 
\!\!\!\!\!
\!\!\!
\upd s 
\vec{N}_{i}^{ext}
.
\end{align}
The resultant particle velocity field \(\phi^{n+1}\vec{v}_{p}^{n+1}\) is
directly enforced on the total velocity field as
\begin{align}
 \vec{v}^{n+1} &= \vec{v}^{**} + 
\left[
\int_{t_{n}}^{t_{n}+h}
\!\!\!\!\!
\!\!\!
\upd s 
\phi
  \vec{f}_{p}
\right],
\\
\left[
 \int_{t_{n}}^{t_{n}+h}
\!\!\!\!\!
\!\!\!
\upd s\phi\vec{f}_{p}
\right]
 &= \phi^{n+1} 
\left(
\vec{v}_{p}^{n+1}-\vec{v}^{**}
\right)-\frac{h}{\rho}\nabla p_{p},
\label{eq:collod_boundary_force}
\end{align}
where \(\phi \vec{f}_{p}\) represents the force density field which
impose the rigidity constraint on the total velocity field.  The
pressure due to the rigidity of the particle is determined by the
incompressibility condition, \(\nabla \cdot \vec{v}^{n+1} = 0\), which
leads to the following Poisson equation for \(p_{p}\), viz, \( \nabla^{2}
p_{p} = \frac{\rho}{h} \nabla \cdot \left[ \phi^{n+1}
\left(\vec{v}_{p}^{n+1}-\vec{v}^{**}\right) \right] \).
We note again that, on the l.h.s. of
Eqs.(\ref{eq:hydro_force}),(\ref{eq:hydro_torque}), and
(\ref{eq:collod_boundary_force}), the integrands \(\vec{F}_{i}^{H}\),
\(\vec{N}_{i}^{H}\), and \(\phi\vec{f}_{p}\) are not explicitly
calculated but their time integrals are solved. In other words, the
solid-fluid interactions are treated in the form of the momentum change,
namely, momentum impulse.

\subsection{Restriction on a time increment}
To enable spatial discretization of the hydrodynamic equations, any standard
scheme, such as the finite difference method, finite volume method, finite
element method, spectral method, lattice Boltzmann discretization and so
forth, can be used. The SP method basically defines a coupling scheme
between the hydrodynamic equations for the solvent and the equations for the discrete colloids.
Since the treatment of the rigidity constraint of
the particle velocity does not introduce an additional time scale, the restriction to a time increment \(h\) is the same as that in uniform fluid cases. This is advantageous as compared to the methods adopted in
refs~\cite{tanaka00:_simul_method_colloid_suspen_hydrod_inter,peskin89:_i},
where a large viscosity or elasticity is used for the velocity over the particle
domain.
For comparison, in the Fluid Particle
Dynamics~(FPD)~method~\cite{tanaka00:_simul_method_colloid_suspen_hydrod_inter},
the large viscosity is 
introduced for the fluid particle \(\eta_{c}\) \((\gg \eta)\) to realize the rigidity constraint.  This means that the
required time increment for FPD should be very small, i.e.,
\(\eta/\eta_{c}(\ll 1)\), as compared with that used in the SP method.

A similar discussion on the restriction of the no-penetration condition in
the advection-diffusion equation of the solutes can be outlined.  One of the
simplest treatments of the no-penetration condition in the particle domain is the penalty method adopted in
refs~\cite{dzubiella03:_deplet_forces_noneq,kodama04:_fluid}, in which an
artificially large potential barrier is introduced for the solutes in the particle domain. The artificial potential should at least be larger
than the other chemical potentials in order to realize no-penetration of the
solutes. This restriction means that the artificial potential requires a smaller
\(h\).  Although this strategy is physically consistent, it is 
numerically inefficient.  In contrast, in the SP method the advection-diffusion requires no additional time scale for the inclusion
of colloids since the no-penetration condition for the solutes is directly
assigned to the solute current in the finite interface domain.
From the above discussion, we can see that the SP method provides us with
much higher numerical efficiency than other methods proposed for direct
numerical simulation of colloidal dispersions.

\section{Results and Discussions}
\subsection{Stokes drag on a periodic array of spheres in a Newtonian fluid}
\begin{figure}
\center
\includegraphics[width=65mm]{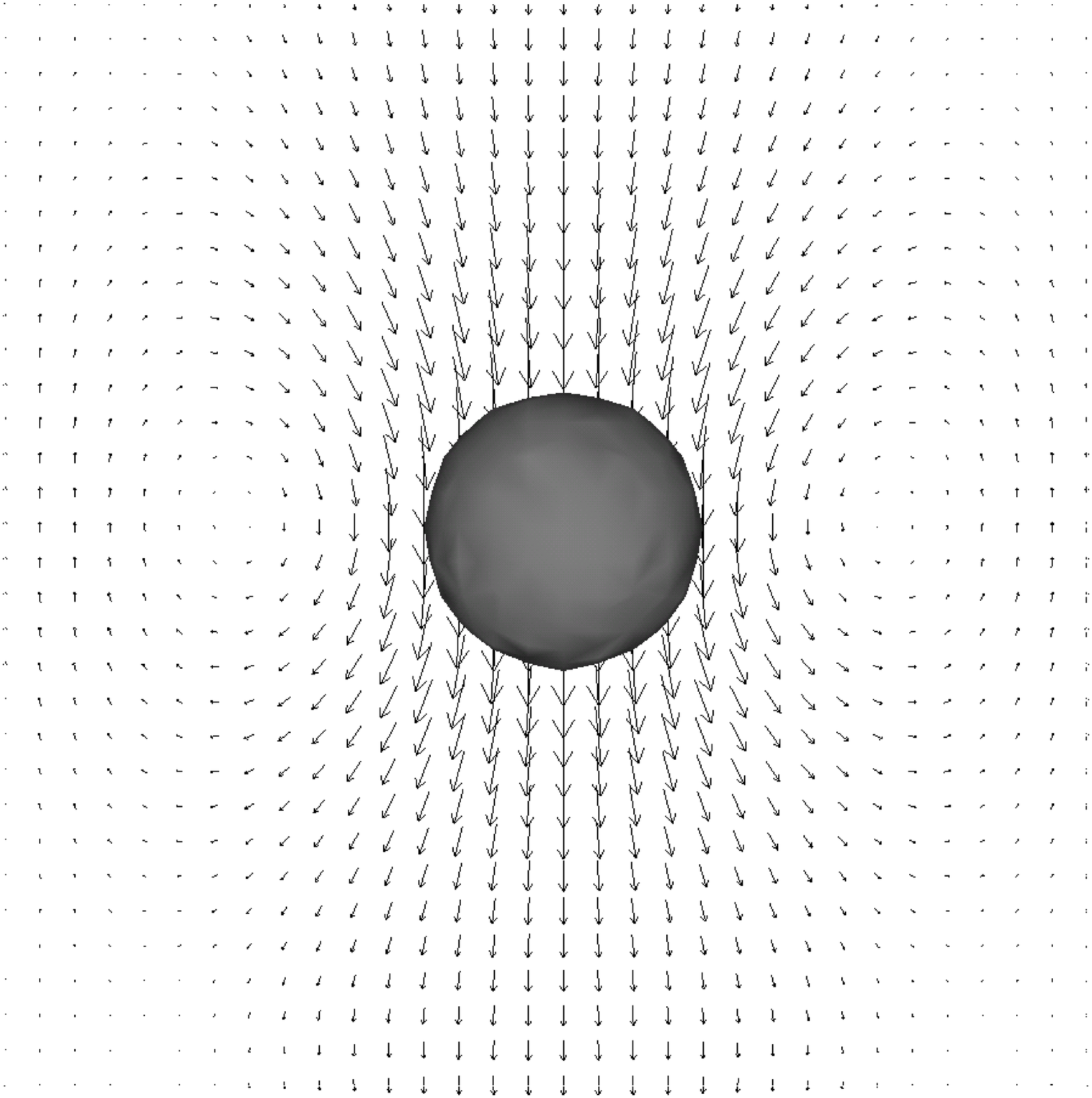}
 \caption{Snapshot of a sedimenting colloid of radius \(a=4\Delta x\).
The arrows indicate the velocity field.}
 \label{fig:sedi_snapshot}
\end{figure}
\begin{figure}
\center
 \includegraphics[width=65mm]{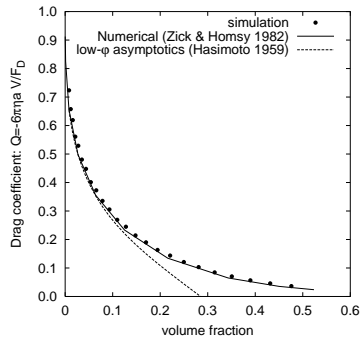} \caption{Drag coefficient of a
 periodic array of spheres in steady Stokes flow as a function of volume
 fraction \(\varphi\) solved using the SP method as compared with the
 analytic result~\cite{zick82:_stokes}~(Solid line) and low-\(\varphi\)
 asymptotics~\cite{hasimoto59:_stokes}~(dashed line).  In this case,
 \(\xi\) is set to unity in the lattice unit.
The range of the volume fraction \(\varphi\) was obtained by changing
the radius \(a\) from \(4\) to \(15.5\) for a lattice size \(\Delta x\)
with \(L=32\Delta x\) fixed.
}  \label{fig:drag}
\end{figure}
To validate the SP method quantitatively, we measured the steady state
drag force on a periodic array of spheres in a Newtonian solvent.  The
velocity distribution around the colloid is depicted in
Fig.\ref{fig:sedi_snapshot}. In general, flow around a colloid occurs as
creep-flow with a Reynolds number of \(Re=aV/\nu \ll 1\).
Figure~\ref{fig:drag} shows the drag coefficient \(Q(\varphi)\), defined
as
\begin{align}
 F_{D} &= -\frac{6\pi\eta a V}{Q(\varphi)},
\end{align}
where \(\varphi = (4/3) \pi (a/L)^{3}\) is the volume fraction in a
cubic box of volume \(L^{3}\).  
 The effect of the boundary condition extends
very far in creep-flow, as the higher the volume fraction, the larger the drag.  Comparison of the drag coefficient given by the SP method with
the analytic result from the Stokes equation by Zick and
Homsy~\cite{zick82:_stokes} verifies the validity of the SP method over the 
entire range of \(\varphi\).

\subsection{Lubrication interaction in a finite system}
One of the most important effects of solvent flow is the lubrication
interaction between nearby particles with relative motions.  The exact
solution of the Stokes equation for two isolated spheres has been
found~\cite{jeffrey84:_calcul_reynol} and has provided much insight into the basic
physics of colloidal suspensions.  However, its application to
many-particle systems through the method of pairwise addition requires care.  For quantitative prediction of the rheology of concentrated suspensions,
numerical results have identified many differences, whether the shear mode of the 
lubrication interaction is included or
not~\cite{ball97:_brown,melrose04:_contin}. There exists a fundamental
ambiguity in the application of the analytic expression of two isolated spheres to a many-particle system in the finite domain using pairwise addition.

\begin{figure}
\center
\includegraphics[width=.8\hsize]{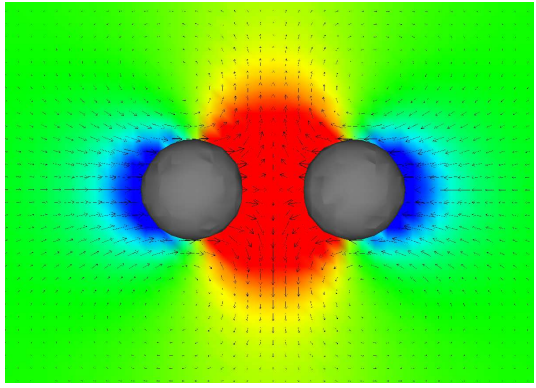}
 \caption{Snapshot of two approaching colloids of radius
 \(a=4\Delta x\).  The arrows indicate the velocity field. The color
 Map around the colloids represents the pressure distribution: a change in
 color from red to blue corresponds to a change from high to low pressure.}
 \label{fig:lub_snapshot}
\end{figure}
\begin{figure}
\center
\includegraphics[width=65mm]{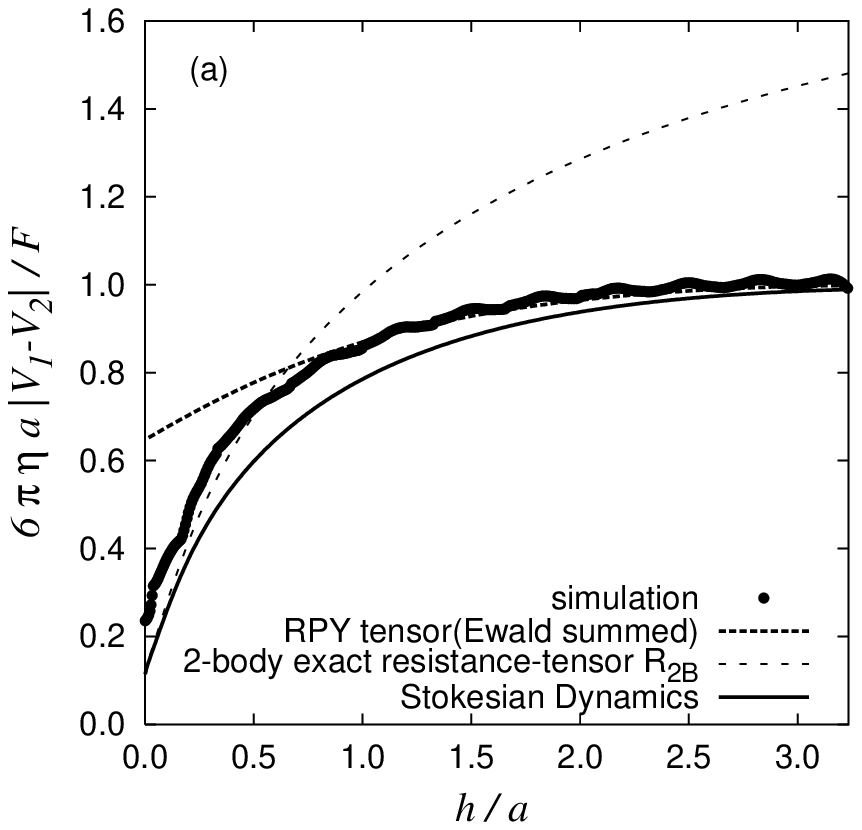}
\includegraphics[width=65mm]{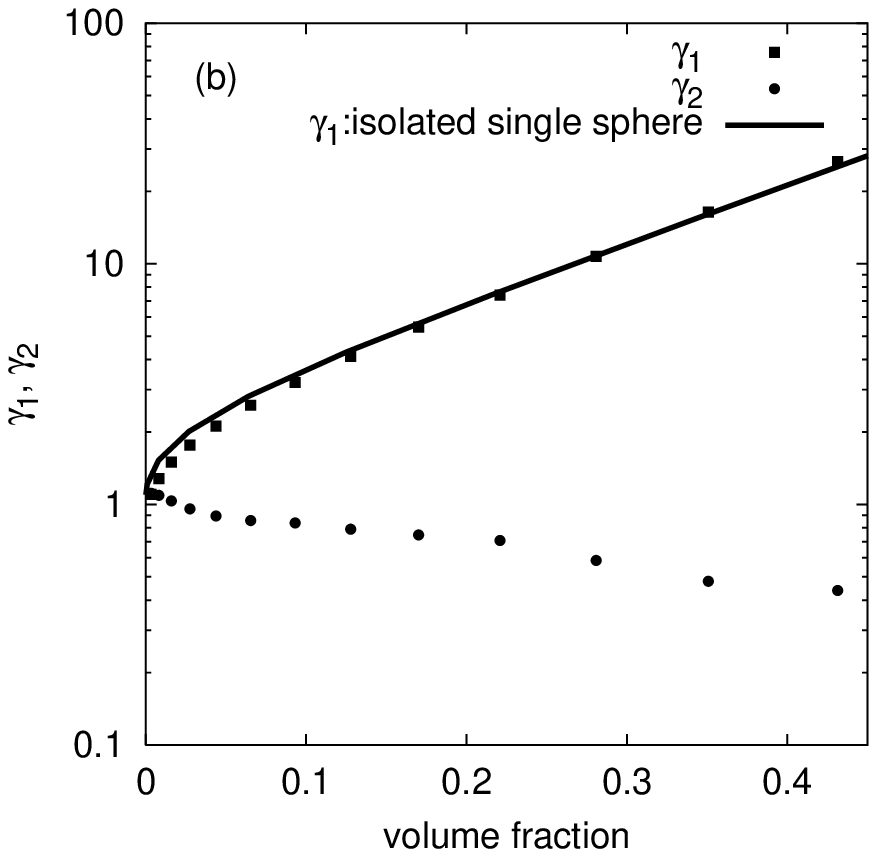}
 \caption{
(a) Relative velocity between two approaching spheres versus the gap between
the sphere surfaces (symbols). The slight oscillations in the SP results are a result of the finite lattice spacing.  In this case, \(\xi\) is set to
unity in the lattice unit.
Theoretical curves are shown for far-field asymptotics using the
Rotne--Prager--Yamakawa tensor~(dotted line)~\cite{beenakker86:_ewald_rotne}, a
near-field exact solution~(dashed line)~\cite{jeffrey84:_calcul_reynol}, and
Stokesian Dynamics~(solid line)\cite{brady88:_stokes_dynam}.
(b) The two coefficients representing the squeeze interaction (scaled to those at infinite dilution) versus volume fraction.  \(\gamma_{1}\)~(square) represents the
one-body drag and \(\gamma_{2}\)~(circle) represents the two-body squeeze
interaction due to relative motion.}
 \label{fig:lub}
\end{figure}

We computed the squeeze lubrication interaction between two approaching
spheres in a finite system.  The velocity and pressure distributions are
depicted in Fig.\ref{fig:lub_snapshot}.  Figure~\ref{fig:lub}(a) shows
the normalized approach velocity of a pair of particles versus the gap
\(h\) between two equal spheres solved by the SP method as compared with
other theories. The two asymptotic solutions at \(h\ll a\) and \(h\gg
a\) are from the exact solution of the isolated
pair~\cite{jeffrey84:_calcul_reynol} and Rotne--Prager--Yamakawa
tensor~\cite{beenakker86:_ewald_rotne}, respectively. The Stokesian
Dynamics~(SD)~solution~\cite{brady88:_stokes_dynam} is based on an
interpolation of these two asymptotic solutions. The simulation result
nicely reproduces not only the two asymptotic regimes but also the
crossover which occurs at \(h/a \sim 0.7\). It is found that SD
underestimates the mobility in this intermediate regime.  This
underestimation is due to the approximation adapted in SD in which the
components are just asymptotic two-body solutions. The result confirms
the relevance of our simulation, thus demonstrating the importance of
the hydrodynamic interaction in a finite system.

We discuss the dependence of the lubrication interaction on the volume
fraction.  By assuming the functional form \(\left|V_{1}-V_{2}\right|/F
= 1/\left(\Gamma_{1}(\varphi)+2\Gamma_{2}(\varphi)/h\right)\) from
lubrication theory, where \(\Gamma_{1}(\varphi)= 6\pi \eta a
\gamma_{1}(\varphi)\) is the one-body drag coefficient and
\(\Gamma_{2}(\varphi)=(3\pi \eta a^{2}/2)\gamma_{2}(\varphi)\) is the
squeeze coefficient, the effect of the volume fraction is represented by
\(\gamma_{1}(\varphi)\) and \(\gamma_{2}(\varphi)\).
These friction coefficients can be extracted by fitting the curve of
Fig.\ref{fig:lub}(a) for each \(\varphi\). 
The reduced coefficients \(\gamma_{1}\) and \(\gamma_{2}\) are plotted in
Fig.\ref{fig:lub}(b) as a function of the volume fraction. The solid line in
Fig.\ref{fig:lub}(b) is from Fig.\ref{fig:drag}.  Although a
periodic array of spheres exhibits different flow geometry from the case of
two approaching spheres, the volume fraction dependence of \(\gamma_{1}\)
of the two cases is comparable.  Moreover, the squeeze coefficient
\(\gamma_{2}\) is found to be a decreasing function of the volume
fraction. In other words, the squeeze mode is most
enhanced at infinite dilution.  In the literature~\cite{kim91:_microh},
it is pointed out that the squeeze coefficient is at least smaller than
that of the exact solution for isolated pairs.  These results further
validate the SP method.

Although the SP method itself is an efficient scheme as a direct
simulation of utility in constructing a more coarse-grained model for suspensions, such as in dissipative particle dynamics, constitutive modeling, etc., the
calculation by direct simulation also gives fundamental information about
the hydrodynamic interactions.

\subsection{Sedimenting charged colloids}
As an example of the specific application of the method to a multi-component fluid, we compute the hydrodynamic drag for sedimenting charged colloids in the absence of an external electric field.  In this case, the sedimenting charged colloids induce a flow that determines a charge distribution which differs from the equilibrium case~\cite{russel89:_colloid_disper,probstein03:_physic_hydrod}.  The skewed ion-distribution gives rise to polarization in the electric double-layer.
Moreover, double-layer deformation induces an electro-osmotic flow, which
makes the flow around a charged colloid different from that of a neutral colloid.

\begin{figure}[htbp]
 \center
 \includegraphics[width=65mm]{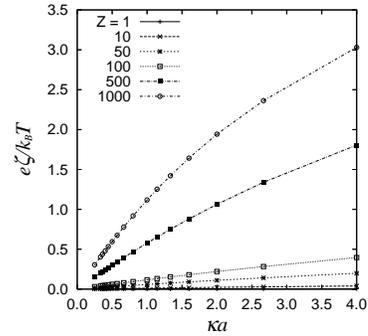}
\caption{Non-dimensionalized zeta potential as a function of
the inverse Debye length \(\kappa a\) and the valence of the colloid
 \(Z\)
computed using the semi-analytic formula of Ohshima--Healy--White
\cite{ohshima82:_accur}. The dotted lines are guides to the eye.}
\label{fig:zeta_dilute}
\end{figure}

In this simulation, a periodic array of charged colloids in a 1:1
electrolyte solution under gravity was computed.  We specify the valence
of the colloids first, and the counterion in the host fluid assures the
charge neutrality of the whole system. The
bulk concentrations of the 1:1 electrolyte \(\bar{C}_{\pm}\) were
determined by specifying the Debye length \(\kappa^{-1} = \left\{4\pi
\lambda_{B}(\bar{C}_{+}+\bar{C}_{-})\right\}^{-1/2} \).
The Bjerrum length \(4\pi\lambda_{B}=e^{2}/\epsilon k_{B}T\) was set to
unity and \(\xi\) was chosen as twice the lattice unit to resolve the
surface charge distribution, which is represented by \(\left|\nabla
\phi\right|\).  For simplicity, the counterion and coion were set to the
Schmidt number \(0.5\), which choice did not affect the qualitative
aspect of the following results but only control the transient time to
the steady state.
The Debye length was chosen so that the effective radius of the
double-layer \(a+\kappa^{-1}\) was smaller than half of the system size
\(L/2\).
In order to clarify the situation in terms of a typical colloid science
parameter, nondimensional zeta potential \(e\zeta/k_{B}T\) is shown in
Fig.\ref{fig:zeta_dilute}, which should be determined by specifying
\(\kappa a\) and the valence of the colloid \(Z\).  The corresponding
dimension of the zeta potential \(\zeta\) for this simulation at
20\(^{\circ}\)C is of the order of \(10\)mV.
We computed the linear response regime for gravitational driving
to observe the effect of electro-osmosis on hydrodynamic
drag.

\begin{figure}
\center
 \includegraphics[width=65mm]{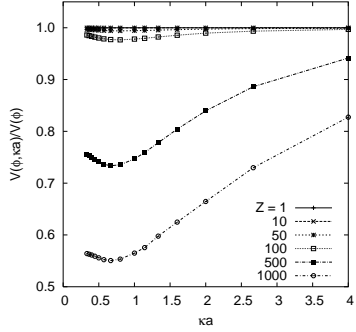}
 \caption{Sedimentation velocity of a periodic array of colloids of
 different valences \(Z\) in electrolyte solutions as a function of the
 inverse Debye length \(\kappa a\).  The ordinate is reduced by the
 sedimentation velocity of neutral colloids at the same volume fraction
 of \(0.008\) (\(a=4\Delta x,~L=32\Delta x\)). The effective volume fraction including the electric
 double-layer defined by \((4/3)\pi \{(a+\kappa^{-1})/L\}^{3}\) was chosen
 to be less than unity.  Dotted lines are guides to the eye.}
 \label{fig:sedi_charge}
\end{figure}

We plot the sedimentation velocity scaled by that of a neutral colloid
as a function of the inverse Debye length scaled by the radius of
colloid in Fig.~\ref{fig:sedi_charge}.  As the valence of the colloid or the
zeta potential increases, the sedimentation velocity reduces to that
of the neutral colloid. The hydrodynamic drag of the
electrolyte solution was most enhanced when the Debye length was
comparable to the size of the colloids.
These
facts qualitatively agree with the analytic result at infinite
dilution~\cite{booth54:_sedim_poten_veloc_solid_spher_partic,ohshima84:_sedim_veloc_poten_dilut_suspen}.

\begin{figure}
\center
\includegraphics[width=.8\hsize]{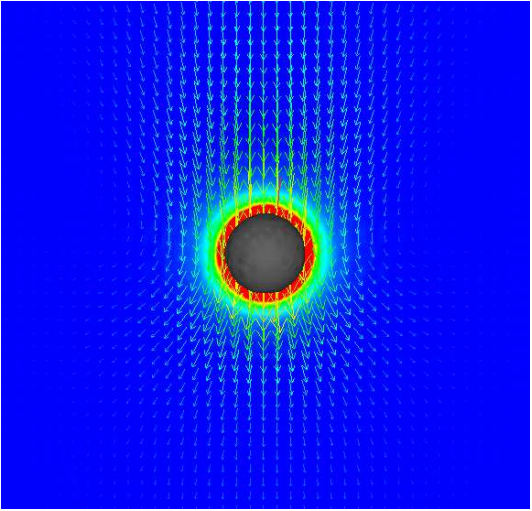}
 \caption{Snapshot of a sedimenting charged colloid when \(\kappa a
 =1\) and \(Z=1000\). The arrows indicate the
 velocity field. The color map around the particle represents the charge
 density field: a change in color from red to blue corresponds to
 a change from high to low charge density.}
 \label{fig:charge_sedi_snapshot}
\end{figure}

The charge-density and the velocity distributions when \(\kappa a=1 \)
and \(Z=1000\) are depicted in Fig.\ref{fig:charge_sedi_snapshot}. In
this case, the charge distribution was largely uniform
and thus the counteracting electrostatic effect of the double-layer
polarization does not have much effect. Therefore, the enhanced
electrohydrodynamic drag can mainly be attributed to the friction
between the solvent and the ions.  Because of this electrohydrodynamic
coupling of the transfer in momentum, the solvent-flow pattern around
the colloid was modified from the case of a neutral colloid. As a
result, the viscous drag on the colloid was enhanced. This mechanism of
enhanced viscous drag by electro-osmotic flow generally exists in
colloids in electrolyte solutions.

It has been known that in an infinitely dilute system (or the thin
double-layer limit) the velocity decays as \(r^{-3}\) in the region of
\(\kappa r\gg 1\) in contrast to the \(r^{-1}\) decay of an infinitely
dilute neutral system, where \(r\) is the radial distance from the center of
the colloid~\cite{anderson89:_colloid_trans_inter_forces,long01}.
However, in the system size adapted in our simulation this asymptotic
regime was not reached.  For \(\kappa r \lesssim 1\), we have the screened
hydrodynamics regime, where the velocity decays as \(v\propto e^{-\kappa
r}/r\)~\cite{long01}.
These factors account for why the flow patterns in Figs.(\ref{fig:sedi_snapshot}) and
(\ref{fig:sedi_charge}) resemble one another.  In other words, the
electrohydrodynamic interaction should be a pronounced effect in small
finite systems as is seen in neutral systems.

We note that, at a finite volume fraction and in the range of the Debye
length, the charged colloid dynamics could be effectively assessed
through the SP method.  Further application of the SP method to
electrophoresis of concentrated suspensions is reported
elsewhere~\cite{kim06:_direc_numer_simul_elect}.

\section{Conclusions}
We have presented a new simulation scheme for colloidal dispersions in a
solvent of a multi-component fluid, which we call the ``Smoothed
Profile~(SP) method''. The SP method improves and extends a previous
method proposed by the
authors~\cite{yamamoto01:_simul_partic_disper_nemat_liquid_cryst_solven,nakayama05:_simul_method_resol_hydrod_inter_colloid_disper,kim04:_effic}.

The description of the colloidal systems is based on the Navier--Stokes
equation for the momentum evolution of the solvent flow, the
advection-diffusion equation for the solute distribution, the rigid body
description of colloidal particles, with dynamical coupling of all these
elements. 
Based on momentum conservation between the continuum solvent fluid
and the discrete rigid colloids, the time-integrated hydrodynamic force and
torque are derived. This
expression of the mechanical coupling between a fluid and particles is
well-suited for numerical simulations in which differential equations are
discretized in time. 
With this formulation relating the solid-fluid interaction, standard
discretization schemes for uniform fluids are utilized as is; no special care is needed for the solid-fluid boundary mesh.
Since the hydrodynamic interaction is solved
through direct simulation of the solvent fluid, the many-body effect can be
fully resolved.

The utility of the SP method was assessed in various test problems,
not only using simple fluids but also simulating charged
colloids in electrolyte solutions. The results confirmed that the SP method
is effective for studying the dynamical behavior of colloids.
Although we have focused on systems of spherical colloids in simple
fluids and Poisson--Nernst--Planck electrolytes (i.e. Poisson--Boltzmann
level description of electrolytes), application to other types of
macromolecules of other shapes, such as disks~\cite{capuani06:_lattic_boltz},
rods, and others, and other constitutive models for solvents is
straightforward.

\end{document}